# Simulation Study of Ge p-type Nanowire Schottky Barrier MOSFETs

Jaehyun Lee and Mincheol Shin

*Abstract*— Ambipolar currents in Germanium p-type nanowire Schottky barrier MOSFETs were calculated fully quantum-mechanically by using the multi-band *k·p* method and the non-equilibrium Green's function approach. We investigated the performance of devices with [100], [110], and [111] channel orientations, respectively, by varying the nanowire width, Schottky barrier height, and EOT. The [111] oriented devices showed the best performance. In comparison to Si as a channel material, Ge is more desirable because more current can be injected into the channel, resulting in steeper subthreshold slope and higher on-state current. Our calculations predict that the Ge channel devices should have an EOT gain of 0.2 ~ 0.5 nm over Si channel devices.

## I. INTRODUCTION

GERMANIUM has recently regained attention as a channel material to replace silicon in metal-oxide semiconductor field-effect transistors (MOSFETs) [1]-[4]. Ge is particularly well suited for p-type devices, thanks to its high hole mobility which is about four times greater than that of silicon, as well as some processing related issues, such as dopant solubility and Fermi level pinning at the valence band.

In this work, we explore the possibility of using Ge as a channel material in p-type MOSFETs with Schottky barrier (SB) contacts or SB-MOSFETs. Our motivation may be summarized as follows. Despite their advantages in reducing short channel effects (SCEs), SB-MOSFETs have the major drawback that on-state drain currents ($I_{ON}$) are limited by the SBs at the channel/silicide interfaces [5]-[12]. A solution to this problem is to use a channel material with low SB height (SBH) and low effective masses to enhance the tunneling current ($I_{tunn}$) in the on-state. In this respect, Ge p-type SB-MOSFETs are quite promising because a low SBH for holes at Ge/metal interfaces is feasible (~ 0 eV or less) and the light hole effective mass ($m_{lh}$) of bulk Ge is as low as 0.046 $m_0$, where $m_0$ is the free electron mass [13]-[15]. Devices with nanowire configuration are considered in this work to assess their ultimate performance limits.

Jaehyun Lee and Mincheol Shin* are with the Department of Electrical Engineering, Korea Advanced Institute of Science and Technology (KAIST), Daejeon, Republic of Korea. (e-mail: mshin@kaist.ac.kr*)

## II. SIMULATION APPROACH

The structure of Ge p-type nanowire SB-MOSFETs is shown in Fig 1. To describe the hole transport in the Ge channel, we adopted the multi-band *k·p* method [16]. The *k·p* parameters for Ge have been adjusted, so that the nanowire subband structures of Ge calculated by the *k·p* method match those by the $sp^3s^*$ tight-binding (TB) method [17].

Simple parabolic effective mass (PEM) Hamiltonians were applied to the metallic source and drain [18-20]. It was also assumed that the virtual valence band top and the effective mass were 1.0 eV above the Fermi level and 0.5 $m_0$ in the transport direction, respectively. These values have little effect on the results in this paper.

Accurate calculation of $I_{tunn}$ is vital in any SB-MOSFET simulations, and in this work, this has been implemented by treating the full quantum transport of holes. That is, the hole density and current were calculated by self-consistently solving Poisson's equation and the non-equilibrium Green's function (NEGF) equation [6] with the 6-band *k·p* Hamiltonian with spin-orbit coupling. Ballistic transport was assumed. The coupled mode-space *k·p* method was employed for efficient calculation [16,21].

Note that electron current was also calculated in this work. Since the band gap of Ge in bulk is only 0.66 eV, one should check whether the off-current ($I_{OFF}$) is limited by the ambipolar current. The electron transport in the n-branch was calculated by considering the eight equivalent Λ valleys of Ge and the generalized effective mass Hamiltonian with the full inverse mass tensor was used.

## III. RESULTS AND DISCUSSION

In our simulations of the Ge p-type nanowire SB-MOSFETs shown in Fig. 1, the Ge channel with a square cross-section of width ($W$) is assumed to be intrinsic. The gates surround the channel, and the channel length ($L$) is scaled as $L = 4W$. The equivalent oxide thickness (EOT) is 1 nm, SBH is 0.2 eV, and the drain voltage ($V_{DD}$) of -0.5 V is applied.

In the following, we first characterize the device performance of Ge p-type nanowire SB-MOSFETs. Fig. 1 shows the drain current ($I_d$)-gate voltage ($V_g$) characteristics of the device of $W$ = 10 nm for the three channel orientations. The currents are ambipolar. Despite the fact that the band gap of Ge is quite small, the electron current in the n-branch has little effect on the off-state hole current in the p-branch. The fact that the effective SBH for electron is considerably raised due to the size quantization effect in nanowires also significantly contributes to the suppression of the electron current. Only the hole current is considered hereafter.

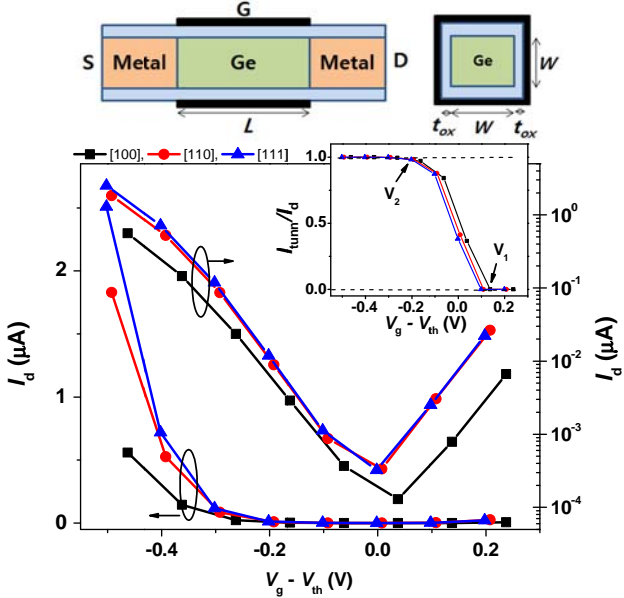

Fig. 1. (Upper panel) A schematic diagram of Ge p-type nanowire SB-MOSFETs. (Lower panel) $I_d$-$V_g$ characteristics of the devices with $W$ = 10 nm and $L$ = 40 nm, displayed in the linear (left axis) and log (right axis) scales. Inset shows the $I_{tunn}$ contribution to $I_d$ in the hole branch.

In the hole branch in Fig. 1, the crossover from the thermionic to the tunneling regimes is seen in the threshold region, which is typical in SB-MOSFETs. That is, in the inset of Fig. 1 where the $I_{tunn}$ contribution to the $I_d$ is shown, the regions of $V_g$-$V_{th}$ > $V_1$ and $V_g$-$V_{th}$ < $V_2$ correspond to the thermionic- and tunneling-current dominant regimes, respectively, and between them, there is a transition region where we measure the subthreshold swing (SS).

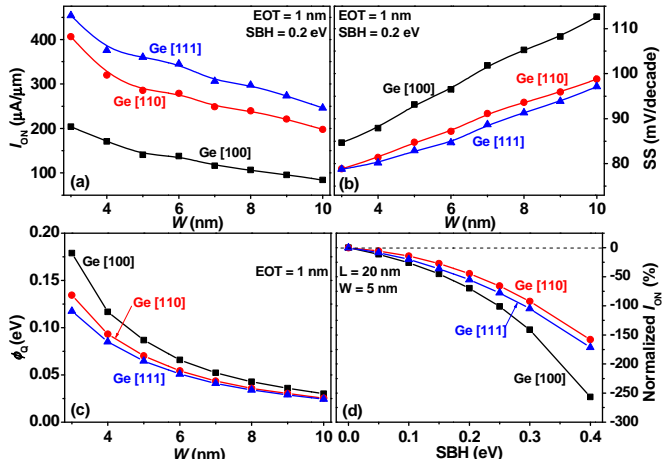

Fig. 2. (a) $I_{ON}$, (b) SS, and (c) $\phi_Q$ of the [100], [110], and [111] channel-orientation devices as $W$ is varied from 3 nm to 10 nm. The device aspect ratio of $L/W$ = 4 is maintained in the graph. (d) $I_{ON}$ (normalized by $I_{ON}$ at SBH = 0 eV) versus SBH for $W$ = 5 nm.

As seen in Fig. 1, the current level of the [111]-oriented device is about 2.5 times larger than that of the [100]-oriented device. The normalized $I_{ON}$ (divided by $W$) versus $W$ in Fig. 2 (a) consistently shows the behavior for the range of $W$ = 3 ~ 10 nm. $I_{ON}$ was calculated with $V_{DD}$ = -0.5 V after adjustment of the gate work function such that $I_{OFF}$ = 10 nA/μm. Note that the increase of $I_{ON}$ with the decrease of $W$ is usually observed in ballistic transistor simulations regardless of whether the contacts are ohmic or Schottky.

The dependence of the SS on $W$ is shown in Fig. 2 (b). As $W$ is reduced, the SS improves, which is due to the fact that current injection is enhanced with improved gate controllability at smaller $W$. The SS in the [111]-oriented devices is steeper than that in the [100]-oriented devices, which can be also clearly seen in Fig. 1. The superiority of the [111] channel orientation with respect to the SS is particularly noteworthy. For the [111] channel orientation, the device with $W$ = 10 nm already achieves a SS of 100 mV/decade or below, whereas $W$ should be reduced to about 6 nm to achieve the same SS for the [100] channel orientation. This 'gain' of 3 ~ 4 nm in $W$ is an attractive factor in favor of the [111] channel orientation.

The performance disparity with respect to channel orientation is largely due to the different $m_{lh}$. The $m_{lh}$ value is lighter in the order of the [100], [110], and [111] directions in bulk, which are 0.046 $m_0$, 0.042 $m_0$, and 0.040 $m_0$, respectively, and the device performance improves in the same order. In fact, the hole effective masses become heavier as $W$ becomes smaller in all the three directions, but the bulk effective masses can still be useful for comparison.

Another important factor that affects $I_{tunn}$ is the size quantization effect. This effectively raises the SBH by the quantization energy ($\phi_Q$), which increases sharply as $W$ is reduced below 5 nm. See Fig. 2 (c). The [100]-oriented device is most disadvantageous. The size quantization effect further lowers the device performance of the [100]-oriented device in comparison with the other orientations. Also, note the fact that the dependence of $I_{ON}$ on SBH is significantly more sensitive in the [100]-oriented device than in the other devices, as shown in Fig. 2 (d), which implies greater variability in orientation.

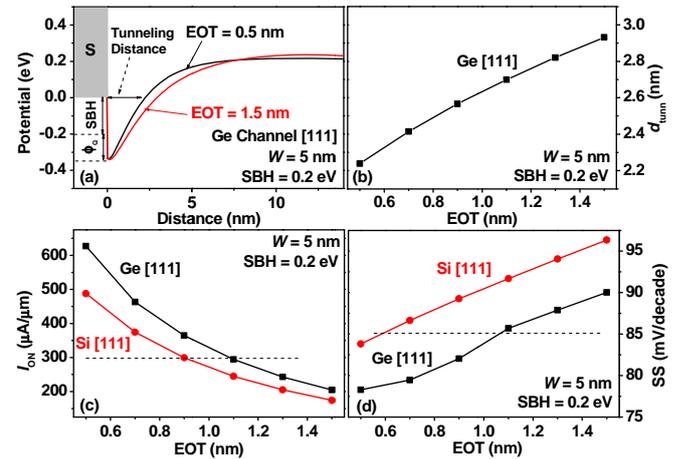

Fig. 3. (a) The on-state potential profiles with EOT = 0.5 and 1.5 nm, respectively. (b) $d_{tunn}$ versus EOT. (c) and (d) compare $I_{ON}$ and SS of Ge channel and Si channel devices, respectively, as EOT is varied from 0.5 nm to 1.5 nm.

Figs. 3 (a) and (b) show the tunneling distance ($d_{tunn}$) in the on-state, which is defined to be the thickness of SB at the source Fermi level. As expected, the shortening of $d_{tunn}$ with the decrease of EOT is observed. Our calculation shows that $d_{tunn}$ is shortened by about 0.7 nm as EOT is reduced by 1.0 nm, leading to a rough estimation that $d_{tunn}$ linearly scales with EOT.

The above results for Ge p-type SB-MOSFETs are qualitatively similar to those obtained when Si is used as the channel material [9]. Quantitatively, however, Ge outperforms Si in all aspects, as will be discussed. The lighter $m_{lh}$ of the former is considered the main factor in the significant difference. For instance, the $m_{lh}$ value of Ge and Si are 0.11 $m_0$ and 0.20 $m_0$, respectively, for nanowire of $W$ = 5 nm [17].

The $I_{ON}$ of Ge channel devices is about 15~30 % larger than that of Si devices. This translates into the EOT gain of 0.2 nm (see Fig. 3 (c)). That is, in achieving the same $I_{ON}$, the EOT of Ge channel devices can be larger by the amount of the EOT gain than that of the Si channel devices. Note that the same SBH of 0.2 eV was used in this work for both Ge and Si channel devices. Considering the fact that the SBH for holes of the former is expected to be lower than that of the latter, the difference in $I_{ON}$ can be further increased. When it comes to SS, there is even larger EOT gain of 0.5 nm as seen in Fig. 3 (d). For instance, to achieve an SS of 85 mV/decade, the EOT of the Si devices should be as small as 0.6 nm, whereas the EOT of the Ge devices can be 1.1 nm. Considering the gate leakage reduction associated with a thicker gate oxide, such an amount of EOT gain is quite significant.

## IV. Conclusion

The device performance of SB-MOSFETs largely depends on how well current can be injected into the channel. In this respect, $m_{lh}$ is the most important factor. Thanks to the small $m_{lh}$ of Ge in the [111] direction, Ge p-type nanowire SB-MOSFETs oriented in this direction show superior performance to the other two orientations considered. In comparison to Si as a channel material, Ge is more desirable. Our calculations predict that Ge channel devices should have an EOT gain of 0.2 ~ 0.5 nm over Si channel devices.

## Acknowledgment

This research was supported by the Pioneer Research Center Program and the Basic Science Research Program thorough the National Research Foundation of Korea (NRF) Funded by the Ministry of Education, Science and Technology (Grant No. 2012-0000459 and No. 2012-0002120)

## References

[1] Jin-Hong Park, Munehiro Tada, Duygu Kuzum, Pawan Kapur, Hyun-Yong Yu, H-.S. Philip Wong, Krishna C. Saraswat, "Low Temperature (≤380 ) and High Perfromance Ge CMOS Technology with Novel Source/Drain by Metal-Induced Dopants Activation and High-K/Metal Gate Stack for Monolithic 3D Integration", in *Proc. IEEE Int. Electron Devices Meeting*, 2008, pp. 1-4.
[2] Huiling Shang, Harald Okom-Schmidt, Kevin K. Chan, Matthew Copel, and John A. Ott, "High Mobility p-channel Germanium MOSFETs with a Thin Ge Oxynitride Gate Dieletric," in *Proc. IEEE Int. Electron Devices Meeting*, 2002, pp. 441-444.
[3] D. S. Yu, C. H. Huang, A. China, C. Zhu, M. F. Le, B. J. Cho, and D. L. Kwong, "Al$_2$O$_3$-Ge-on-insulator n- and p-MOSFETs with filly NiSi and NiGe dual gates," *IEEE Electron Device Lett.*, vol. 25, no. 3, pp. 138-140, Mar. 2004.
[4] Rui Zhang, Takashi Iwasaki, Noriyuki Taoka, Mitsuru Takenaka, and Shinichi Takagi, "High-Mobility Ge pMOSFET With 1-nm EOT Al$_2$O$_3$/GeO$_x$/Ge Gate Stack Fabricated by Plasma Post Oxidation," *IEEE Trans. Electron Devices*, vol. 59, no. 2, pp. 335-341, Feb. 2012.
[5] John M. Larson and John P. Snyder, "Overview and Status of Metal S/D Schottky-Barrier MOSFET Technology," *IEEE Trans. Electron Devices*, vol. 53, no. 5, pp. 1048–1058, May 2006.
[6] Jing Guo and Mark S. Lundstrom, "A Computational Study of Thin-Body, Double-Gate, Schottky Barrier MOSFETs," *IEEE Trans. Electron Devices*, vol. 49, no. 11, pp. 1897-1902, Nov. 2002.
[7] Mincheol Shin, "Computational Study on the Performance of Multiple-Gate Nanowire Schottky-Barrier MOSFETs," *IEEE Trans. Electron Devices,* vol. 55, no. 3, pp. 737-742, 2008.
[8] Moongyu Jang, Yarkyeon Kim, Jaeheon Shin, Seongjae Lee, and Kyoungwan Park, "A 50-nm-gate-length erbium-silicided n-type Schottky barrier metal-oxide-semiconductor field-effect transistor," *Appl. Phys. Lett.*, vol. 84, no. 5, pp. 741-743, Feb. 2004.
[9] Mincheol Shin, "Effect of channel orientation in p-type nanowire Schottky barrier metal-oxide-semiconductor field-effect transistors," *Appl. Phys. Lett.*, vol. 97, 092108, Sep. 2010.
[10] Chi-yui Ahn and Mincheol Shin, "Ballistic Quantum Transport in Nano-scale Schottky Barrier Tunnel Transistors," *IEEE Trans. Nanotechnology,* vol. 5, no. 3, pp. 278-283, 2006.
[11] Mincheol Shin, Moongyu Jang, and Seongjae Lee, "Quantum mechanical simulation of charge distribution in Schottky barrier MOSFETs," *J. Kor. Phys. Soc,*. vol. 45, pp. S547-S550, 2004.
[12] Mincheol Shin, Moongyu Jang, and Seongjae Lee, "Quantum Simulation of Resonant Tunneling in Nanoscale Tunnel Transistors," *J. Appl. Phys.*, vol. 99, mo. 6, pp. 066109, 2006.
[13] D. Z. Chi, H. B. Yao, S. L. Liew, C. C. Tan, C. T. Chua, K. C. Chua, R. Li, and S. J. Lee, "Schottky barrier height in germanide/Ge contacts and its engineering thorugh germanidation induced dopant segregation," in *Proc. IEEE IWJT*, 2007, S5-2, pp. 81-85.
[14] Shiyang Zhu, Rui Li, S. J. Lee, M. F. Li, Anyan Du, Jagar Singh, Chunxiang Zhu, Albert Chin, and D. L. Kwong, "Germanium pMOSFETs With Schottky-Barrier Germanide S/D, High-κ Gate Dielectric and Metal Gate," *IEEE Electron Device Lett.*, vol. 26, no. 2, pp. 81-83, Feb. 2005.
[15] Tatsuro Maeda, Keiji Ikeda, Shu Nakaharai, Tsutomu Tezuka, Naoharu Sugiyama, Yoshihiko Moriyama, and Shinichi Takagi, "High Mobility Ge-on-Insulator p-Channel MOSFETs Using Pt Germanide Schottky Source/Drain," *IEEE Electron Device Lett.*, vol. 26, no. 2, pp. 102–104, Feb. 2005.
[16] Mincheol Shin, "Full-quantum simulation of hole transport and band-to-band tunneling in nanowires using the *k·p* method," *J. Appl. Phys.*, vol. 106, 054505, Sep. 2009.
[17] Mincheol Shin, Sunhee Lee, and Gerhard Klimeck, "Computational Study on the Performance of Si Nanowire pMOSFETs Based on the *k·p* Method," *IEEE Trans. Electron Devices*, vol. 57, no. 9, pp. 2274-2283, Sep. 2010.
[18] Mincheol Shin, "Three-dimensional quantum simulation of multigate nanowire field effect transistors," *Mathematics and Computers in Simulation*, vol. 79, no. 4, pp. 1060-1070, 2008.
[19] Mincheol Shin,."Efficient simulation of silicon nanowire field effect transistors and their scaling behavior," *J. Appl. Phys.* vol. 101, no. 2, pp. 024510, 2007.
[20] Mincheol Shin ,"Quantum Simulation of Device Characteristics of Silicon Nanowire FETs,", *IEEE Trans. on Nanotechnology*, vol. 6. no. 2, pp. 230-237, 2007.
[21] Mincheol Shin, "Quantum transport of holes in 1D, 2D, and 3D devices: the k ·p method," *J. Computational Electronics*, vol 10, pp. 44-50, Feb. 2011.